# High power fast pulsed magnetron discharges

M. Ganciu

Laboratoire de Physique des Gaz et des Plasmas, UMR CNRS-UPS 8578,
Université Paris-Sud, Bat 210, 91405 Orsay Cedex, France
*E-mail : Mihai.Ganciu@pgp.u-psud.fr*

*In order to improve the quality of thin film deposition by magnetron discharges, particularly by an effective ionization of the sputtered vapor, we developed an ionized physical vapor deposition method based on pre-ionized pulsed magnetron discharges operating in microsecond range (1-50 µs) leading to a lower breakdown delay (~1 µs), short fall time of the pulsed magnetron current (<1 µs), a high ion-to-neutral flux ratio at the substrate, an arc free regime and good stability with very large instantaneously power. The transition to the self-sputtering regime, the reactive operation mode and the ion transport to the substrate can be modulated by the current pulse shape and duration. High cathode current densities (1-20 $A/cm^2$) are obtained and large pressure range operation (3-300 mTorr) is demonstrated.*

## Introduction

Magnetron sputtering has become a major technique in the high-rate deposition of technologically important thin films and coatings [1]. Amongst difficulties arising during the deposition of highly insulating films such as oxides and nitrides is the generation of local arcs on the cathode or substrate by poisoning the surface in DC magnetron sputtering due to the ion charging on the surface of the dielectric film. A modulation of the DC magnetron discharge with frequencies in the range of 10–100 kHz is a possible method for the compensation of the ion surface charging by an electron flux to the surface of dielectric formed layer. Moreover, modulating the discharge also appears to have some benefit for the thin film properties in ion-assisted film growth due to the changes in the intrinsic plasma properties by increasing the time-averaged electron temperature and density at the substrate compared with the DC equivalent plasma. Kelly et al. [1] have shown that pulsing the plasma in the frequency range 20-100 kHz during the reactive sputtering of alumina can give significant improvements in film density, hardness, stoichiometry and optical properties, compared to DC sputtered films. This was explained by higher ion flux emerging to the substrate in the mid frequency pulsed DC magnetron operation.

These results can motivate the increasing interest in the last years for ionized physical vapor deposition (IPVD) techniques (more ions than neutrals in the deposition flux) to improve the film quality as well as the deposition on complex surfaces needed in industrial applications [2-4]. In these reactors, the sputtered species, mainly emitted by the cathode as neutral particles, are ionized in the area between the target and the substrate by additional plasma of several types: inductively coupled plasmas, electron cyclotron resonance plasma, microwave discharge plasma and hollow cathode magnetron plasma. Despite good results concerning vapor ionization, there are some difficulties concerning internal antennas subjected to sputtering or vapor deposition. Moreover in the reactive sputtering, due of the different ionization ratio of atomic or molecular species, problems arise for trench and via filling [3]

A promising alternative technique was developed [5-7] using high-power pulses applied directly on the magnetron cathode. For a short time (some tens of microseconds) the high target current density (several $A/cm^2$) sustains very dense plasma near the cathode which induces efficient sputtered vapor ionization. Nevertheless, this method is limited by a too large delay for avalanche development, a relatively high probability of arc formation and a lower deposition rate in comparison to DC magnetron sputtering systems [8].

In this work, we describe a novel technique of magnetron sputtering by using a fast pulsed-magnetron discharge operating in a preionization regime [9 - 18]. This technique allows faster current growth, high ion-to-neutral flux ratio at the substrate, arc-free regime, and good stability with very large instantaneous power. At the same time, the average power is comparable to that obtained with currently used pulsed magnetrons. The arc development is greatly reduced when using short pulses if the pulse duration is lower than the time necessary for an emission site to become overheated. In the reactive operation mode, the duration of the current pulse is limited by the need to prevent the breakdown of the positively charged dielectric film. For example, in the case of a cathode current density of about 1 $A/cm^2$, one obtains a pulse duration limit of about 3 µs.

**Experimental**

The developed experimental devices (Figure 1) include classical magnetrons with rectangular or circular cathodes made from different metals: titanium, copper, ruthenium etc. The power supply and discharge plasma were considered as a coupled system [19]. An adapted pulse generator was used to apply high power pulses (1-50 µs, 10-200 A, 500-1200 V, 0-1 kHz) directly on the cathode and insuring short fall time (<1 µs) of magnetron current pulses [9 - 18]. It allows also the superposition of a DC, RF or microwave additional pre-ionization discharge. The microwave and RF discharges can be induced in the region between the target and the substrate by microwave applicators [4, 20] or by a coil antenna [21]. In the case of RF pre-ionization, emission spectroscopy measurements near the substrate can be performed. The time resolved emission spectroscopy measurements used in this study was performed by using a THR 1500 Jobin-Yvon monochromator equipped with a Hamamatsu PM connected through an adapted 50 Ω coaxial cable on a 100 MHz digital Tektronix oscilloscope. Voltage and current measurements were provided by classical large bandwidth probes. The ionic and electronic currents on the substrate were measured by using a biasing system which allows 10 ns time resolution.

**Results and discussion**

The system maintains a preionization magnetron current at least at ~4 mA in between the pulses. Corresponding self-established cathode voltage was ~ -300 V. The presence of the initial preionization plasma ensures the fast and reproducible cathode current dynamics when high voltage is applied.

A typical pulsed applied voltage on the cathode (target) and magnetron current for argon working gas at 1.4 Pa and DC preionization of 10mA are presented in Figure 2 In a pressure range of 0.4 -5 Pa and for different cathode geometries we obtained maximum current density of 1 to 20 $A/cm^2$.

For a titanium target a typical optical emission spectrum near the cathode is presented in Figure 3a. With RF preionization we have been able to make emission spectroscopy measurements near the substrate (Figure 3b). For cathode current densities higher than 1$A/cm^2$, we observed an enhancement of Ti II lines. Moreover we also observed this enhancement near the substrate.

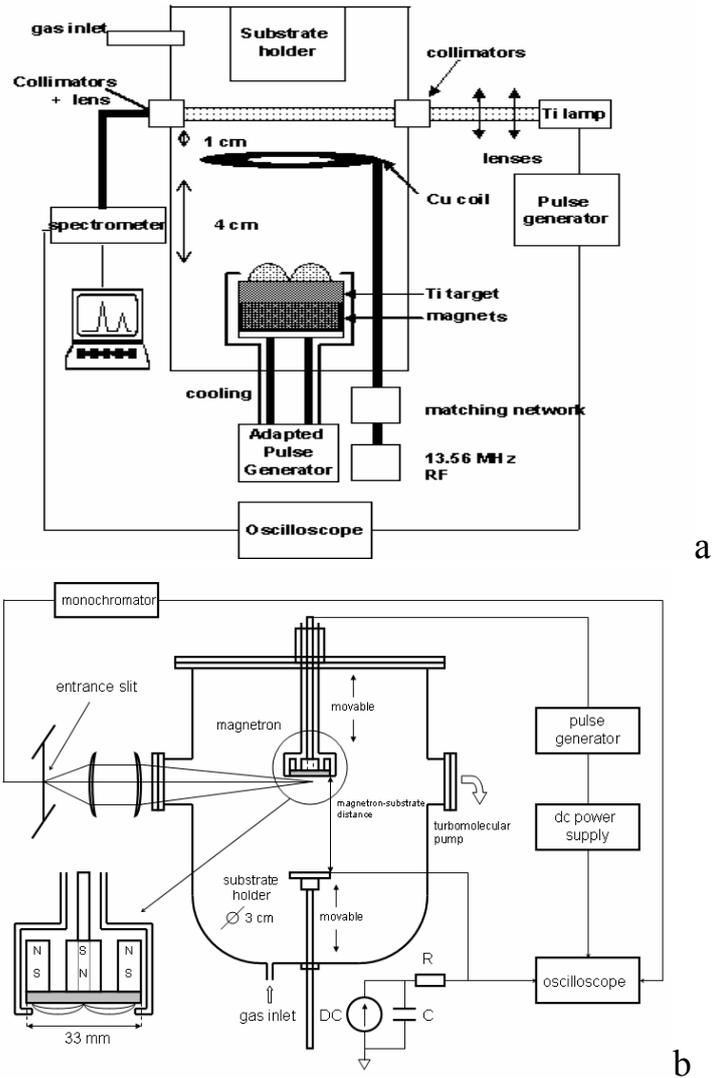

*Figure 1: Two developed pulsed magnetron experimental devices: a) For IPVD of titanium [9, 10, 14, 15] and b) for IPVD and self-sputtering of copper [13]*

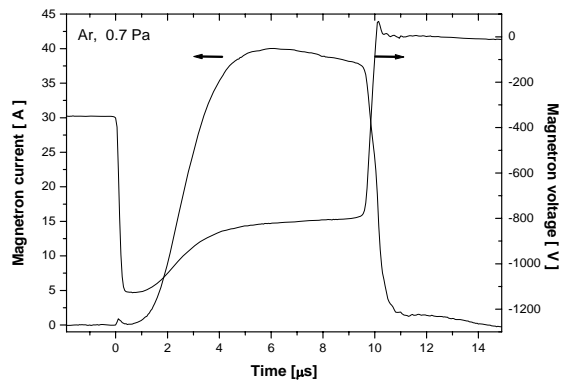

*Figure 2 : Typical cathode voltage and current waveforms*

By pulsing a conventional titanium hollow cathode lamp [20, 21] synchronously with magnetron current pulses it was possible to implement time resolved atomic spectroscopy to measure both Ti and $Ti^+$ concentration and to estimate the vapor ionization ratio. Thus, for a Ti magnetron cathode we estimated the Ti vapor degree of ionization near the substrate to be ~ 70% [11, 15].

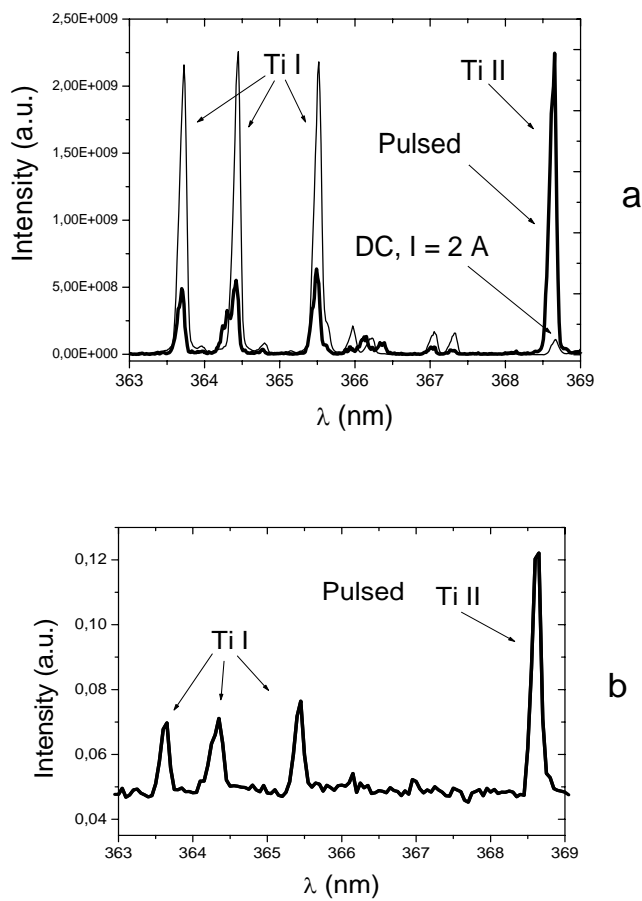

*Figure 3: Optical emission spectrum near the target (a) and near the substrate (b)*

The pulsed ion current has been measured on the substrate biased at -40 V and for different magnetron discharge configuration and pre-ionization conditions. Figure 4 presents the time evolution of the ion current on the substrate for different RF powers recorded using the experimental device presented in Figure 1a. An efficient gas ionization between cathode and substrate can optimize the ionized vapour. This ionization can be obtained in different ways (DC, RF, microwave, multi-pulse operation)

Proper pre-ionization allows reducing the breakdown delay to an acceptable range to work in very short pulse regime, such that electric arcs are avoided. Thus, a stable operation is obtained at high frequency pulse repetition rate (~ 1 kHz) and high current densities (1-20 $A/cm^2$). The pulse duration can be adapted for each magnetron configuration to obtain maximum ion-to-neutral ratio without electric arc development. In reactive mode operation, the optimum pulse duration was in the range of 1-10 µs [11, 17] and for non-reactive mode it was in the range of 5 – 30 µs [11-15].

Due to the high ionization degree of sputtered vapor, this method allowed, for some metals to achieve during the pulse a transition to sustained self-sputtering regime [13], with advantages in reduced buffer gas scattering and its inclusion in deposited films. The improved power coupling of our system allowed fast magnetron current rise when high voltage is applied and a better control of transition towards the self-sputtering regime within few $\mu$s.

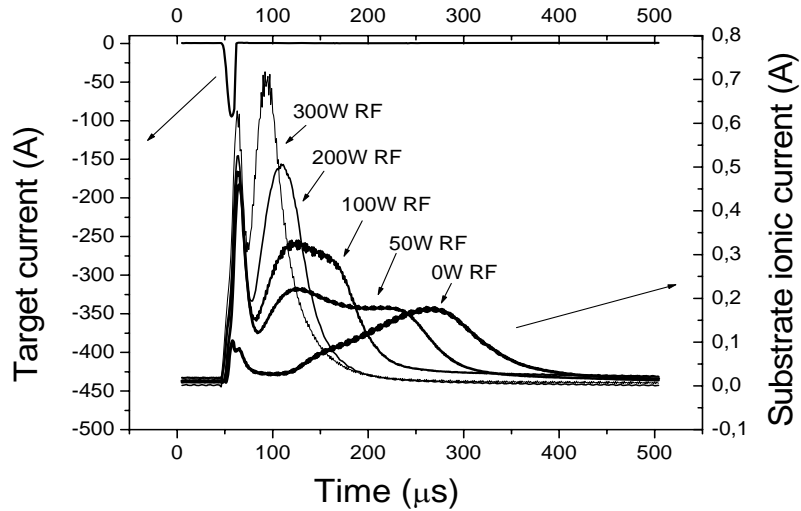

*Figure 4: Target current and ion current on the substrate for different RF power; target to substrate distance of 8cm*

We investigated these phenomena by using the experimental device presented in Figure 1b. [13]. The time evolution of the investigated copper and argon lines is presented in Figure 5. We can identify a transition between buffer gas discharge and a discharge sustained in metallic vapors. This transition is associated with the falling of Ar I lines investigated in magnetron plasma simultaneously with a transition to stable Cu II emissions. After the transition to the metallic vapor sustained discharge, we can assume that we achieved a stable self-sputtering regime. After the transition to self-sputtering of copper cathode by $Cu^+$ ions, a stationary state for Cu I lines is achieved when neutral Cu atoms fill the whole volume of the magnetized region where atoms are excited by electrons of the glow. Due to the system facility to ensure fast fall time of the pulsed magnetron current, one can stop the current just in the moment when self-sputtering begin to be important. Thus by changing the pulse duration, from 1 $\mu$s up to about 50 $\mu$s, it is possible to choose the proportion of the self - sputtering regime in the overall sputtering process.

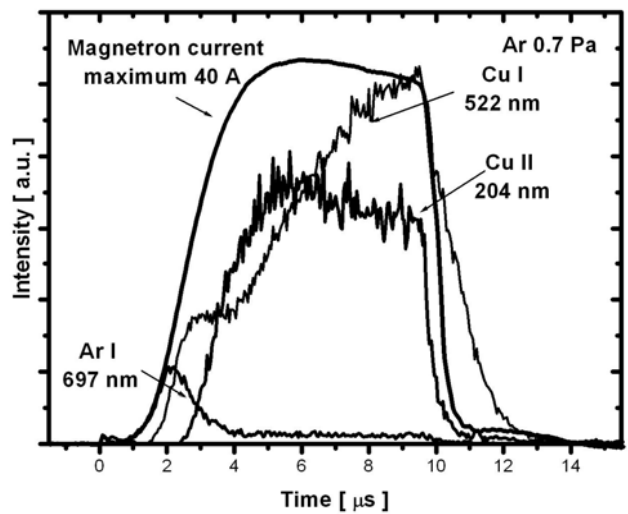

*Figure 5: Time evolution of pulsed magnetron current and argon and copper line intensities.*

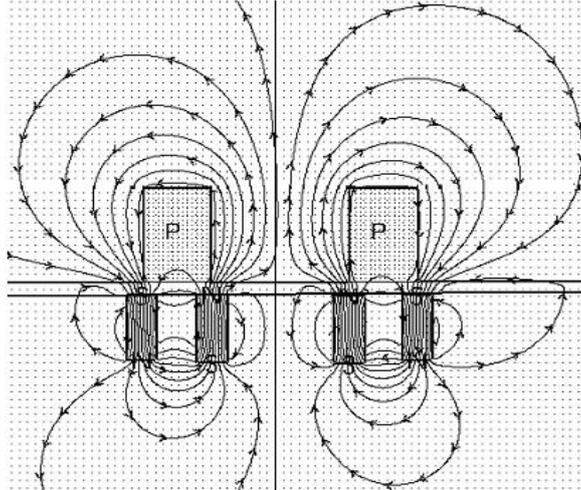

*Figure 6: Simulation of magnetic field perturbation by the presence of high density plasma (P). In this simulation for the diamagnetic effect of the dense plasma we assume plasma relative magnetic permeability of 0.1. The magnetic field in the absence of the plasma is shown in the bottom of the figure.*

For high current cathode densities, the cathode plasma kinetic pressure can be higher than the magnetic pressure. Taking into account only the electron contribution $n_e kT_e$ and assuming electronic temperature ~ 3 eV, a simple estimation for B = 200 gauss and J = 4A/cm$^2$ gives kinetic pressure higher than magnetic one. The electron density was estimated from ion density current by using the Bohm model [22]. Thus we can suppose that at high cathode current densities, which implies also high electronic densities we can favor plasma expansion by magnetic plasma deconfinement [10, 23]. Moreover as the ions in self sputtering mode result from vapor ionization, we can assume that the ion temperature is given by the average energy of sputtered vapor (~ 3 eV). If the plasma kinetic pressure is much larger than the magnetic pressure, in general the plasma pushes the magnetic field around and carries it along with its natural motions. An example of a simple qualitative simulation of magnetic field perturbation induced by the cathode dense plasma in expansion due of diamagnetic properties [10] is presented in Figure 6.

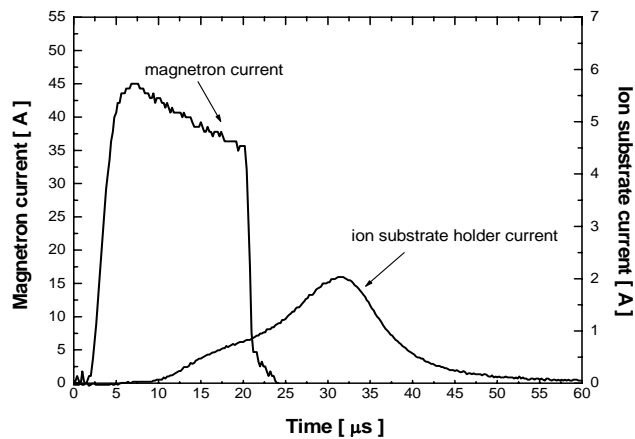

*Figure 7: Pulsed cathode current and ion substrate current shapes*

In high power DC magnetron discharges for self sputtering operation, Posadowski attributed the plasma de-confinement to the magnetic field induced by Hall current [24] but in the fast pulsed regime of magnetron current this effect is reduced by induced Foucault current in the metallic cathode.

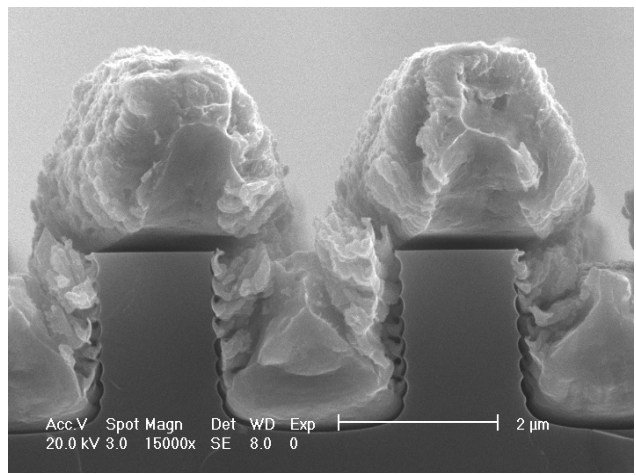

*Figure 8: Copper trench filling for 1.3 Pa argon pressure, target current density of 10A/cm$^2$, 20 µs pulse duration, 100 Hz repetition frequency and average power on the target of about 10W/cm$^2$ Distance from magnetron to substrate was 6 cm and substrate bias was -40V.*

For a pulsed magnetron current and an ion current on the substrate presented in Figure 7, we obtained a well-filled trench structure of aspect ratio 1 by a compact copper deposit as can be seen in Figure 8 [10, 13]

Transporting the ionized vapor to the substrate holder is not straightforward; however, efficient transport and collection of the ionized vapor ensure convenient deposition rates. A key aspect is to ensure the charge compensation of the ion beam originated from ionized vapor ejected from the cathode with relatively high kinetic energy (1-10 eV). The transport of ions originating from sputtered vapor to the substrate can be improved if these ions could be replaced by background gas ions in order to ensure space charge neutralization of the magnetized plasma. Both criteria can be fulfilled by creating an additional plasma between the cathode and the substrate. We analyzed these phenomena by adding a secondary inductively coupled plasma [14, 16] and simple diffusion models for sputtered particle transport were developed [16]. These diffusion models are valuable only for relatively high working pressure (> 4 Pa) when mean free path of sputtered particles is much lower than the characteristic distance between cathode and substrate and are limited by the buffer gas rarefaction during long pulses. An experimental analysis by optical emission spectroscopy of the complex transport of sputtered particles in a high power pulsed magnetron operating at argon pressure of 4 Pa with titanium cathode has been done in [12] The experimental results was explained supposing initial ballistic transport of the sputtered particles followed by simultaneous thermalization and scattering of sputtered particles by buffer gas to all directions.

The influence of pulse duration on the titanium vapor deposition rate in fast high-power magnetron discharges has been investigated by using an experimental device showed in Figure 1 [15]. Deposition rate has been recorded for different effective powers in DC magnetron discharges (mdc) and in high-power magnetron discharge modes (HPPM) for argon pressure of 2 and 10 mTorr. In HPPM regime, for the same effective power, but for different repetition frequencies, the deposition rate increases as the pulse length decreases. At 2 mTorr, 300 W of effective power, the deposition rate achieved with pulses 5 µs long is ~70% of the one obtained with the mdc discharge. At the same pressure, same power, but with longer pulses (20 µs), the deposition rate in HPPM is ~20% of the one measured in mdc. In contrary, the metallic vapor ionization rate, as determined by absorption measurements, diminishes as the pulses are shortened. Nevertheless, the ionization rate is in the range of 50% for 5 µs pulses while it is below 10% in the case of a classical continuous magnetron discharge. The reduction of the deposition rate for longer pulses was mainly attributed to two physical phenomena. First, the self-sputtering passage with low self-sputtering yield of titanium atoms (lower to unity for applied voltage pulse characteristics) could be invoked. Second, metal ions must be able to diffuse freely towards the substrate and the space charge of the ion flux must be compensated. This can be achieved by ambipolar diffusion [16, 25] but with some difficulties connected to the magnetic plasma confinement or by supplementary plasma generated between cathode and substrate [14, 25].

Another way to rise the deposition rate, by improving the recuperation of the ionized vapor accumulated in magnetized plasma, is to shorten the fall time of the magnetron current and to apply an inverse pulse on the cathode. Therefore, the accumulated ions in magnetized plasma are directed to the substrate. Preliminary results on a compact table-top pulsed magnetron, developed for large pressure range operation [18], showed an ion current rising just after the magnetron current turn-off, as observed on Figure 9. The experimental device is quite similar to that presented in Figure 1b and with an external access on one side to the substrate holder. The compact configuration of the device allows a control of the plasma confinement with external magnets near the substrate.

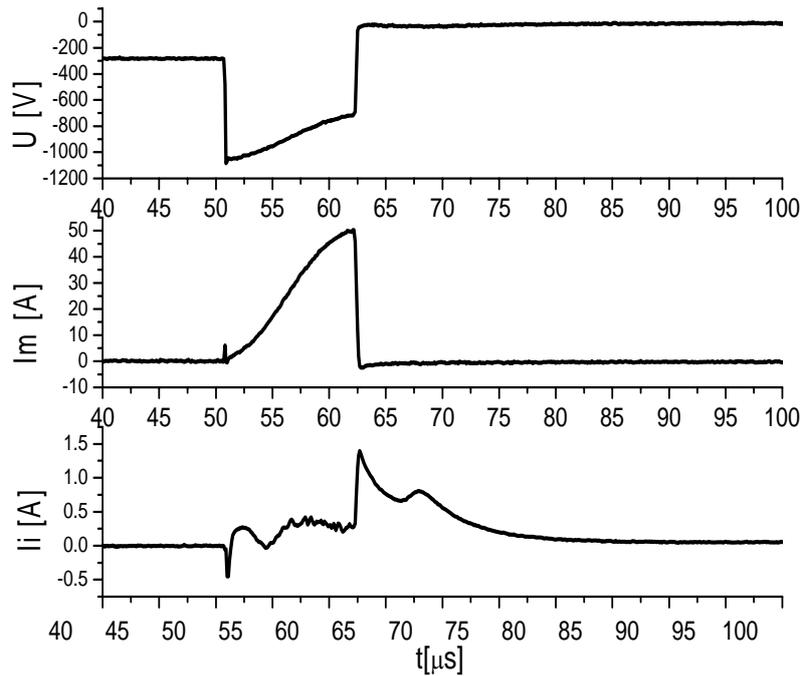

*Figure 9: Cathode voltage (U), magnetron current (Im) and substrate ionic current (Ii) for 300 mTorr Ar pressure, 3 cm diameter of the copper cathode, 1.5 cm cathode-substrate distance, and -40 V substrate bias voltage*

This relatively high pressure processing, which works with standard pumping conditions, still yields a high ion to neutral particle fluxes on the substrate. The pulsed operation induces gas rarefaction at the cathode improving the vapour transport even at relatively high pressure (~ 300 mTorr). By ensuring the thermalization of energetic particles through neutral-neutral and ion-neutral collisions, it permits an easier control of the power transferred to the depositing film compatible with low thermal budget operation. In particular, we guess to reduce the destructive effect on the substrate of the fast neutrals reflected on the cathode [26]. For reactive regime operation the additional gas ionization by fast neutrals should be important to compensate the differences of ionization ratios of atomic or molecular species [3] needed for growing thin layers

A slightly more elevated pressure operation ensures supplementary vacuum pumping by buffer gas flow but also brings additional benefits towards its industrial application by significantly decreasing the vacuum pumping requirements and significantly reducing the cost of operation.

Moreover the same principles, by transposition and scaling, will permit to move on in the diagnostic of thin layers by controlled pulsed sputtering, hence optimizing the classical analytical methods GD-OES (glow discharge optical emission spectroscopy) and GD-MS (glow discharge mass spectroscopy) [27].

**Conclusions**

In conclusion we present a fast high-power pulsed-magnetron IPVD technique which uses pre-ionization to significantly reduce the breakdown delay to the acceptable range needed to work in very short pulse regime, such that electric arcs are avoided. Thus, stable operation are obtained at high frequency pulse repetition rate (~ 1 kHz) and high current densities (1-20 A/cm$^2$), for a large pressure range buffer gas operation (3-300 mTorr). Efficient gas initial ionization between cathode and substrate can optimize ionized vapor collection. This pre-ionization can be obtained in different ways (DC, RF, microwave, multi pulse operation). The current pulse duration and shape can be adapted for every magnetron configuration, in order to obtain maximum ion-to-neutral ratios without electric arc development, to control the transition towards the self-sputtering regime, and to optimize the reactive mode operation. The first results of trench filling using the described fast pulsed IPVD process make it promising for industrial thin-film deposition. We report preliminary results for a copper cathode pulsed magnetron discharge operated at relatively high pressure (~300 mTorr), with different pulse parameters and magnetic field distributions for the deposition of thin film on polymer substrates.